\documentclass[a4paper,12pt,reqno]{amsart}
\usepackage{fullpage}
\usepackage{amssymb}
\usepackage{amsthm}
\usepackage{mathtools}
\usepackage{graphicx}
\usepackage{tikz}
\usepackage{tikz-cd}

\usepackage[titletoc]{appendix}

\newcounter{counter2}
\numberwithin{equation}{section}
\numberwithin{counter2}{section}

\newtheorem{thm}[counter2]{Theorem}
\newtheorem{prop}[counter2]{Proposition}

\newtheorem{rem}[counter2]{Remark}
\newtheorem{exmp}[counter2]{Example}

\newcommand{\coordindex}[1]{\mbox{\normalsize\ensuremath{\textstyle #1}}}

\DeclareSymbolFont{script}{U}{eus}{m}{n}
\DeclareMathSymbol{\Wedge}{0}{script}{"5E}

\newcounter{mnotecount}[section]

\renewcommand{\themnotecount}{\thesection.\arabic{mnotecount}}

\newcommand{\mnote}[1]
{\protect{\stepcounter{mnotecount}}$^{\mbox{\footnotesize
$
\bullet$\themnotecount}}$ \marginpar{
\raggedright\tiny\em
$\!\!\!\!\!\!\,\bullet$\themnotecount: #1} }

\def\be{\begin{equation}}

\def\ee{\end{equation}}

\def\bea{\begin{eqnarray}}
\def\eea{\end{eqnarray}}

\begin{document}
\title{Heavenly equations in de Sitter space II: \\ Solutions and twistor theory}

\author{Timothy Moy$^1$}
\address{University of Sheffield \\ Hicks Building,  Hounsfield Rd,  S3 7RH}
\email{t.moy@sheffield.ac.uk} 
\date{\today}
\thanks{${}^1$London Mathematical Society Early Career Research Fellow,  School of Mathematical and Physical Sciences,  University of Sheffield}

\begin{abstract}
Given an anti-self-dual (ASD) Einstein metric in Przanowski's normal form,  we obtain the metric in Lipstein-Nagy (LN) normal form.  This allows us to find simple,  rational solutions of the LN equation corresponding to the analytically continued Fubini-Study metric on $\mathbb{CP}^2$,  the ASD Taub-NUT metric with cosmological constant,  and the one-loop universal hypermultiplet metric.  Next,  for a metric in LN form,  we provide formulae for the contact structure and associated Jacobi bracket on twistor space.  This is the structure that gives rise to $\Lambda$-deformed celestial symmetry algebras that have been the subject of recent interest.  Finally,  we study the linearised LN equation via a version of Dunajski-Mason recursion. In particular,  we find a contour integral formula to obtain solutions to this linearisation from contact Hamiltonians on twistor space.  In the case of ASD Taub-NUT with cosmological constant,   we explicitly integrate the twistor distribution and specify Hamiltonians in terms of  elementary expressions in the LN coordinates pulled back to correspondence space.  
\end{abstract}

\maketitle
 \setcounter{tocdepth}{1}
\tableofcontents

In \cite{DM26} it was demonstrated that every complex analytic anti-self-dual (ASD) Einstein metric with non-zero cosmological constant $\Lambda$
could locally be put into a form depending on a single scalar potential $W$ satisfying a second-order non-linear partial differential equation (PDE) introduced by Lipstein and Nagy \cite{LN}.  This PDE closely resembles and may be thought of as a $\Lambda \ne 0$ analogue of Pleba\'nski's second heavenly equation \cite{Pl76}.  It is preceded by similar equations,  in \cite{FinPleb,  CP},  in this regard.  \par 
Much earlier,  Przanowski \cite{PrzanKill,PrzanHerm} obtained a different normal form for ASD Einstein metrics with non-zero $\Lambda$.  The associated PDE,  also a second-order non-linear equation for a single scalar potential,  has been studied in detail in,  for example, \cite{APV,Hoeg}.  Whereas the Lipstein-Nagy (LN) normal form is adapted to a choice of self-dual foliation,  the Przanowski normal form is adapted to a choice of integrable complex structure.  The associated equation more closely resembles Pleba\'nski's first heavenly equation.  
\par 
A question posed in \cite{LN} and left unanswered in \cite{DM26} was the relationship between the two normal forms.  Here,  we provide a procedure for passing between the two.  The idea is that,  given a metric in Przanowski form,  the (say) $+i$-eigenspace of the complex structure distinguished by the coordinate system provides a natural choice of self-dual foliation.  To this one may adapt a coordinate system in which the metric takes Lipstein-Nagy form.  \par 
The details of the construction allow,  in many cases,  explicit integration.  For example,  we write the (analytic continuation of the) Fubini-Study metric in Lipstein-Nagy form.  We also consider the ASD Taub-NUT metric with cosmological constant,  studied in,  for example,  \cite{GP78, N17, BH26}.   For this example,  we in fact first put the metric into the normal form of Tod,  where the metric is determined by a solution of the $SU(\infty)$-Toda equation.   \par 
Another loose end in \cite{DM26} that we tie up here is explicitly realising the Jacobi bracket of contact Hamiltonians on twistor space.
This structure has been the subject of interest in celestial holography and beyond,  and gives rise to what may be referred to as $\Lambda$-deformed celestial symmetries or $\Lambda$-deformed $Lw_{1+\infty}$ algebras.  See,  for example,  \cite{TZ24, BBHKMS24,  BH26,  SW26}.    \par 
More broadly,  the motivation for studying the LN equation is in the multitude of powerful applications of its $\Lambda = 0$ antecedent,  especially in recent years.  The second heavenly equation has been employed in studying gravitational scattering amplitudes:  \cite{Miller1} develops a perturbiner expansion for the second heavenly equation which is used to derive a new proof of the maximally helicity violating formula for gravitons.  \cite{BHS} studies scattering on an Eguchi-Hanson space by using that metric's second heavenly form.  The second heavenly form of important ASD metrics has enabled surprisingly explicit realisations of their non-linear twistor correspondences: In the case of ASD Taub-NUT such realisations \cite{Kim25,AAS26} have been used in the recent twistor construction of the Schwarzschild spacetime \cite{AASS26}.   The twistor incidence relations of the ASD Pleba\'nski-Demia\'nski family ($\Lambda = 0$)  are obtained in terms of elementary functions of the spacetime coordinates and twistor parameter in \cite{Kim26}.   
Pleba\'nski's second heavenly equation also makes transparent the double copy structure of anti-self-dual gravity.  Mathematically,  this is related to the fact that Pleba\'nski's second heavenly equation is transparently the zero curvature condition for a connection with infinite-dimensional gauge group the Poisson automorphisms of a two-dimensional complex symplectic surface.  Indeed,  anti-self-dual gravity can be viewed as a symmetry reduction of anti-self-dual Yang-Mills with this gauge group \cite{Str95}. The LN equation exhibits a similar structure as remarked upon in \cite{LN,KL25}.  
Finally,  a more far-removed application is the appearance \cite{BS21} of the second heavenly equation and its generalisations in the programme to geometrise Donaldson-Thomas theory via Joyce structures.  
\subsection*{Acknowledgements} 
This article grew out of discussions during the Simons Programme on Twistor Theory and its Applications at IMPAN Warsaw in 2026.  
I would like to thank Tim Adamo,  Bernardo Araneda,  Roland Bittleston,  Maciej Dunajski,  James Lucietti,  Noah Miller and Paul Tod for interesting and helpful conversations. \par 
The research was supported by the Simons Foundation
grant  SFI-MPS-T-Institutes-00010825,  and by the State Treasury funds as part of a
task commissioned by the Minister of Science and Higher Education under the project Organization of the Simons Semesters at the Banach Center - New Energies in 2026-2028 (MNiSW/2025/DAP/491).  
\subsection*{AI disclosure} 
ChatGPT 6 Astra was used at various stages of the research process. The author wrote the article according to their own understanding.  
\subsection*{Notation and conventions}
We use the Einstein-summation convention throughout.  We employ spinor indices following the conventions of \cite{PR84,PR86}, with values $A \in \{0,1\}$ and $\epsilon^{AB},  \epsilon_{AB}$ the standard symplectic matrices satisfying $\epsilon_{AC}\epsilon^{BC} = \delta_{A}^{B}$ with $\epsilon_{01} = 1$.   We raise and lower indices using these objects without comment.  So,  for example, 
$W^{A} := \epsilon^{AB}W_{B}$.  We use the same conventions for primed spinor indices $A'$.  \par 
Let $X$ be an oriented four-dimensional complex manifold throughout.  Given a complex analytic metric $g$,  a $2$-form may be uniquely decomposed into parts that are self-dual (SD) and anti-self-dual (ASD) under the Hodge star operator.  We say $g$ is ASD if the Weyl tensor,  viewed as an endomorphism on $2$-forms,   sends ASD 2-forms to ASD 2-forms and is zero otherwise.  \par We say $g$ is Einstein if the Ricci tensor is proportional to the metric and take the convention that $\mathrm{Ricci}(g) = 6 \Lambda g$ for such a metric.  
\section{The Lipstein-Nagy equation}\label{review}
 The following is shown in \cite{DM26}:
\begin{thm}\label{DM26theorem}
Given a complex analytic ASD Einstein metric $g$,  there exist local coordinates $(z^{A},  \theta^{A})$ for $X$ such that the metric takes the form
\begin{align}\label{general_integrable}
g = \frac{2}{\phi^2}(d \theta_{A} dz^{A} + (\partial_{\coordindex{\theta^B}} W_{A})dz^{A} dz^{B}),
\end{align}
where
\begin{align}
\phi &= \Lambda \theta^1 + z^0,  \label{conf_factor} \\
W_{0} &=\phi\,\partial_{\coordindex{\theta^0}} W,  \quad W_{1} =\phi\,\partial_{\coordindex{\theta^1}} W - 3\Lambda W  \label{W_potential}
\end{align}
for some function\footnote{As a word of warning,  the function $W$ here is the function $\tilde{W}$ in \cite{DM26}.} $W(z,\theta)$,  and $\Lambda := R/24$,  $R$ being the scalar curvature.  Furthermore,  $W$ satisfies the Lipstein-Nagy equation $\text{\small{$\Psi[W]$}} = 0$,  where
\begin{equation}
\begin{aligned}\label{LN}
\text{\small{$\Psi[W]$}} := \epsilon^{AB}(\partial_{\coordindex{\theta^A}}\partial_{\coordindex{z^B}} W 
-\frac{\phi}{2}\epsilon^{CD} (\partial_{\coordindex{\theta^A}}\partial_{\coordindex{\theta^C}} W)\partial_{\coordindex{\theta^B}}\partial_{\coordindex{\theta^D}} W 
-2 \Lambda(\partial_{\coordindex{\theta^A}} W)\partial_{\coordindex{\theta^B}}\partial_{\coordindex{\theta^0}} W). 
\end{aligned}
\end{equation}
  \par 
Conversely,  any such metric \textup{(\ref{general_integrable})} with $W$ satisfying $\Psi[W] = 0$ is ASD and Einstein with scalar curvature $R$. 
\end{thm} 
If $W$ and the coordinates are taken to be real,  the metric signature is split (also known as \textit{Kleinian}).  The theorem allows for $\Lambda = 0$,  in which case an elementary coordinate transformation recovers Pleba\'nski's second heavenly equation (see \cite{DM26} \S 4.2).  \par
Note that the leaves of constant $z^{A}$ define a self-dual foliation for $X$ (that is,  a foliation defined by a decomposable self-dual $2$-form).  In fact,  we may choose to adapt our coordinates to realise any self-dual foliation of $X$ this way.  
One notable aspect of the ansatz (\ref{general_integrable}) is that there is an explicit family of scalar-flat conformally related metrics: those of the form $\phi^2 f g$ for some non-vanishing $f(z^0,z^1)$.  This is a feature of the ansatz and does not require $W_{A}$ satisfy any PDE\footnote{The anti-self-duality condition for a metric of the form (\ref{general_integrable}) (so not assuming (\ref{W_potential})) reduces to a pair of third-order PDEs on $W_{A}$.  The metric (\ref{general_integrable}) is a general ansatz for metrics that are one-sided Petrov type III (or more algebraically special).  See \cite{DFK} for details.}.  \par 
We now introduce notation for a convenient tetrad:
\begin{align}
\label{Evectors}
\mathbf{E}_{A0'} = \partial_{\coordindex{\theta^A}},\quad
\mathbf{E}_{A1'} = \partial_{\coordindex{z^A}} + \epsilon^{BC}(\partial_{\coordindex{\theta^A}} W_{B})\partial_{\coordindex{\theta^C}},
\end{align}
which gives a null-tetrad satisfying $g(\mathbf{E}_{AA'},  \mathbf{E}_{BB'}) = \phi^{-2}\epsilon_{AB}\epsilon_{A'B'}$.  Note that it is \textit{not} normalised as usual.  Directly from the definition of the tetrad we have 
 \begin{align}
 [\mathbf{E}_{A0'},\mathbf{E}_{B1'}] =  -(\partial_{\coordindex{\theta^A}}\partial_{\coordindex{\theta^B}} W^{C})\mathbf{E}_{C0'}, \quad 
 [\mathbf{E}_{A1'},\mathbf{E}_{B1'}] = -2 \mathcal{Q}( W_{[A}) \mathbf{E}_{B]0'}\label{brac2} ,  
 \end{align}
where we have defined the second-order differential operator $\mathcal{Q} :=  \mathbf{E}_{11'}\mathbf{E}_{00'}-\mathbf{E}_{01'}\mathbf{E}_{10'}$,  which up to scale has the same principal symbol as the wave operator.  \par 
\section{Anti-self-duality and the non-linear graviton}\label{twistors}
In this section we introduce the requisite twistor theory and make it concrete in terms of the data of \S\ref{review}.  \par 
Firstly,  the choice of tetrad $\mathbf{E}_{AA'}$ defines a local isomorphism 
\begin{align}\label{paraconformal}
TX \cong S \otimes S',  
\end{align} where $S$ and $S'$ are rank two vector bundles equipped with skew-forms $\epsilon,  \epsilon'$ that satisfy $\epsilon \otimes \epsilon' = \phi^2 g =: \tilde{g}$.  We pick trivialisations $\{e_{A}\},  \{e_{A'}\}$,  $A = 0,1$ for the bundles $S$ and $S'$ so that $e_A \otimes e_{A'}$ is identified with $\mathbf{E}_{AA'}$ under (\ref{paraconformal}) and the skew-forms have components $\epsilon_{AB}$ and $\epsilon_{A'B'}$ respectively.  We denote fibre coordinates on the bundle $S'$ by $\pi^{A'}$.  We use $\lambda = -\pi^{0'}/\pi^{1'}$ for the fibre coordinate on the projectivisation and identify $\mathbb{P}S' \cong X \times \mathbb{CP}^1$.  \par 
Anti-self-duality is equivalent to the existence of functions $l_{A}$ on $X \times \mathbb{CP}^1$ such that $[\mathbf{L}_0,\mathbf{L}_1] = 0$,  where
\begin{align}
\mathbf{L}_{A} = \mathbf{E}_{A1'} - \lambda \mathbf{E}_{A0'} - l_{A} \partial_{\coordindex{\lambda}}.  
\end{align}
$L := \text{span}\{\mathbf{L}_0,\mathbf{L}_1\}$ is the \textit{twistor distribution} on $X \times \mathbb{CP}^1$.  In particular,  anti-self-duality implies the existence of the \textit{twistor space} $Z = (X \times \mathbb{CP}^1) / L$.  We may work in a neighbourhood of any point such that each $\alpha$-surface is simply connected and such that $Z$ is a well-behaved three-dimensional complex manifold (the leaf space of $L$).  Each point $x \in X$ defines a \textit{twistor line} $\mathbb{CP}^1 \hookrightarrow Z$ corresponding to the $\alpha$-surfaces containing $x$.  
Each $\alpha$-surface $\Sigma$ defines a section $\sigma$ of $S'$ over $\Sigma$ up to scaling by a holomorphic function: $T\Sigma$ consists of tangent vectors of the form $\mu \otimes \sigma$ with respect to the identification (\ref{paraconformal}).  \par  
Computing the commutator using (\ref{Evectors}) and (\ref{brac2}) we see that anti-self-duality implies $l_{A} =\mathcal{Q}(W_{A})$.  
Now  (4.1) in \cite{DM26},  which follows by direct calculation from the metric ansatz without enforcing any PDE,  is:
\begin{align}
\mathcal{Q}(W_{A})
&=
\phi \partial_{\coordindex{\theta^A}} \Psi[W] - 3\Lambda \delta_{A}^{\, 1}\Psi[W] + \partial_{\coordindex{\theta^A}} \beta,
\end{align}
where
\begin{align}
\beta := 
-\partial_{\coordindex{\theta^1}} W
-\Lambda \partial_{\coordindex{z^0}} W
-\Lambda^2(\partial_{\coordindex{\theta^0}} W)^2,
\end{align}
and we recall $\Psi[W] = 0$ is the LN equation.   
In particular,  assuming the LN equation we have $l_{A} = \partial_{\coordindex{\theta^A}} \beta$.  \par 
The Levi-Civita connection for $\tilde{g}$ is induced by unique connections on $S$,  $S'$ preserving $\epsilon,  \epsilon'$.  For the connection on $S'$ we compute the connection coefficients $\tilde{\Gamma}_{AA'B'}{}^{C'}$ for the scalar-flat metric $\tilde{g} = \phi^2 g$ with respect to the trivialisation $\{e_{A'}\}$.  The only non-vanishing components are (see also \cite{MoyThesis} \S 5): 
\begin{align}\label{spin_connection}
\tilde\Gamma_{A0'1'}{}^{0'}
&=-\frac12
\partial_{\coordindex{\theta^A}}\partial_{\coordindex{\theta^B}} W^B, \quad 
\tilde\Gamma_{A1'0'}{}^{0'}
=\frac14
\partial_{\coordindex{\theta^A}}\partial_{\coordindex{\theta^B}} W^B,\nonumber \\
\tilde\Gamma_{A1'1'}{}^{1'}
&=-\frac14
\partial_{\coordindex{\theta^A}}\partial_{\coordindex{\theta^B}} W^B,  \quad 
\tilde\Gamma_{A1'1'}{}^{0'}
=-\mathcal{Q}(W_A) = -\partial_{\coordindex{\theta^A}} \beta, 
\end{align}
where we have used $\Psi[W] = 0$ for the last equality.  \par 
Later,  we will need to consider the line bundles $\mathcal{O}(k)$,  $k \in \mathbb{Z}$,  on twistor space (see \cite{MW} for full details). Briefly,  recall that on an $\alpha$-surface $\Sigma$,  there is a line subbundle $L_\Sigma$ of $S'$ locally trivialised by any choice of defining spinor $\sigma$,  and that in the ASD setting,  the Levi-Civita connection induces a flat connection $\nabla^\Sigma$ on this bundle.  The bundle $\mathcal{O}(-1)$ may then be defined as the bundle on $Z$ with fibre at a point (that point corresponding to an $\alpha$-surface $\Sigma$) that is the vector space of parallel sections of the induced connection on $L_\Sigma$.  This condition is conformally invariant.  The reason for the notation is that $\mathcal{O}(-1)$ restricts to the usual tautological bundles on twistor lines $\mathbb{CP}^1 \hookrightarrow Z$ corresponding to points in $X$.   Finally,  $\mathcal{O}(k)$ may be obtained the obvious way from the operations of tensor power or inverse.   \par 
In practice,   the condition for a function $\psi$ on $X \times \mathbb{CP}^1$ to define a section of $\mathcal{O}(k)$ is:
\begin{align}\label{ok_section_practical}
\mathbf{L}_A(\psi)
=
k\Lambda 
(\partial_{\coordindex{\theta^A}}\partial_{\coordindex{\theta^0}} W) \psi.  
\end{align}
For $k = -1$,  evaluating (\ref{ok_section_practical}) on an $\alpha$-surface $\Sigma$ and using (\ref{spin_connection}) shows that $\psi$ represents a parallel section of $\nabla^\Sigma$ with respect to the trivialisation $\sigma = e_{1'}-\lambda e_{0'}$ for $L_\Sigma$.  \par 
The $\Lambda \ne 0$ analogue of Penrose's non-linear graviton \cite{Penrose} is due to Ward \cite{Ward}: there is a one-to-one correspondence between holomorphic contact structures on the twistor space $Z$,  transverse to twistor lines,  and conformally compatible Einstein metrics with non-zero cosmological constant (the metric being defined up to an overall constant scale).   \par 
Let $\mathbf{F}^{AA'}$ be the dual-tetrad defined by
\begin{align}
 \mathbf{F}^{BB'}(\mathbf{E}_{AA'}) = \delta_{A}^{B}\delta_{A'}^{B'}.  
\end{align}
Explicitly:
\begin{align}
\mathbf{F}^{A0'}=d\theta^A+(\partial_{\coordindex{\theta^B}} W^A)\,dz^B, \quad 
\mathbf{F}^{A1'}=dz^A.
\end{align}
Now letting $\Gamma_{AA'B'}{}^{C'}$ be the primed spin-connection coefficients associated to $g$ and the normalised tetrad $\phi \mathbf{E}_{AA'}$,  the conformal transformation rule gives
 \begin{align}
\Gamma_{AA'B'}{}^{C'}
=
\phi\tilde{\Gamma}_{AA'B'}{}^{C'}
-\delta_{A'}^{C'}\mathbf{E}_{AB'}(\phi)
+\frac12\delta_{B'}^{C'}\mathbf{E}_{AA'}(\phi).
\end{align}
In general,  the contact structure on $Z$ associated to an Einstein scale may be defined by the following $1$-form $\tau$ on $X \times \mathbb{CP}^1$:
\begin{align}
\tau=\phi d\lambda -
\left(
\Gamma_{AA'1'}{}^{0'}
-2\lambda\,\Gamma_{AA'0'}{}^{0'}
-\lambda^2\Gamma_{AA'0'}{}^{1'}
\right) \mathbf{F}^{AA'}.  
\end{align}

By construction\footnote{See \cite{MW} Chapter 13 for the proof that this satisfies the required properties.  We have multiplied the $1$-form by $\phi$ for convenience.},  this satisfies $d\tau \wedge \tau \ne 0$,  annihilates $L$,  descends to a $1$-form on $Z$ (which,  abusing notation,  we shall also denote by $\tau$) with values in the line bundle $\mathcal{O}(2)$,  and has kernel transverse to twistor lines $\mathbb{CP}^1 \hookrightarrow Z$ corresponding to points in $X$.   In any case,  its kernel descends to a well-defined contact structure on $Z$.  \par 
Calculating this in terms of the Lipstein-Nagy data,  the $1$-form on $X \times \mathbb{CP}^1$ is 
\begin{align}
\tau
=\phi d\lambda+\phi(\partial_{\coordindex{\theta^A}} \beta)\mathbf{F}^{A1'}
+\Bigg(
\delta_{A}^{\,0}+\phi^2\partial_{\coordindex{\theta^A}}
\Bigg[
\frac{\lambda-\Lambda \partial_{\coordindex{\theta^0}} W}{\phi}
\Bigg]
\Bigg)
(\, \mathbf{F}^{A0'}+\lambda \mathbf{F}^{A1'}).
\end{align}
It is evident that this annihilates $L$,  and furthermore,  that the kernel descends implies $\mathcal{L}_{\mathbf{L}_A}\tau = h_{A}\tau$ for some functions $h_{A}$.  To obtain $h_{A}$  we calculate
\begin{align}
(\mathcal L_{\mathbf{L}_A}\tau)\big(\partial_{\coordindex{\lambda}}\big) &= \mathbf{L}_{A}(\phi)-\tau\big(\big[\mathbf{L}_A,\partial_{\coordindex{\lambda}} \big]\big) 
=\delta_{A}^{\,0} -\Lambda\lambda \delta_{A}^{\, 1}
+\Lambda \partial_{\coordindex{\theta^A}} W_0 - \tau(\mathbf{E}_{A0'}),  
\end{align} 
where we have used $\big[\mathbf{L}_{A},\partial_{\coordindex{\lambda}}\big] = \mathbf{E}_{A0'}$.  Hence
\begin{align}\label{tau_descent}
\mathcal L_{\mathbf{L}_A}\tau
=
2\Lambda
(\partial_{\coordindex{\theta^A}}\partial_{\coordindex{\theta^0}} W)\,\tau.
\end{align}
In particular,  $\tau$ is normalised in order that if a vector field $\mathbf{V}$ on $X \times \mathbb{CP}^1$ descends to a vector field on $Z$,  $\tau(\mathbf{V})$ is a function on $X \times \mathbb{CP}^1$ representing a section of $\mathcal{O}(2)$.  
\section{Przanowski to Lipstein-Nagy}\label{PtoLN}
In this section,  we start from a metric written in Przanowski form and provide a coordinate transformation that puts it in Lipstein-Nagy form.  We will not need the twistor theory introduced in \S \ref{twistors} in this section.   \par 
The following normal form is due to Przanowski:
\begin{thm}[Przanowski normal form \cite{PrzanKill}]
Given a complex analytic ASD Einstein metric with $\Lambda \ne 0$,  there exist local \textup{(}complex\textup{)} coordinates $(u^{A}, \tilde{u}^{A})$ such that the metric takes the form
\begin{align}\label{Prz_metric}
g= 
(\partial_{\coordindex{u^A}}\partial_{\coordindex{\tilde{u}^B}} K)\,du^A d\tilde{u}^B
-\frac{2}{\Lambda}e^{-\Lambda K}\,du^1d\tilde{u}^1,
\end{align}
where $K(u^{A}, \tilde{u}^{A})$  satisfies
\begin{multline}\label{Prz_eqn}
\frac12\,\epsilon^{AB}\epsilon^{CD}
\bigl(\partial_{\coordindex{u^A}}\partial_{\coordindex{\tilde{u}^C}} K\bigr)
\bigl(\partial_{\coordindex{u^B}}\partial_{\coordindex{\tilde{u}^D}} K\bigr)
-\left(\frac{2}{\Lambda}\partial_{\coordindex{u^0}}\partial_{\coordindex{\tilde{u}^0}} K+(\partial_{\coordindex{u^0}} K)(\partial_{\coordindex{\tilde{u}^0}} K)\right)e^{-\Lambda K}=0.  
\end{multline}
Conversely,   every such metric is ASD and Einstein with scalar curvature $R = 24\Lambda$.  
\end{thm}With respect to the coordinate tetrad of $(u^{A}, \tilde{u}^{A})$,  the metric (\ref{Prz_metric}) has determinant $e^{-2\Lambda K}\bigl[(\partial_{\coordindex{u^0}} K)(\partial_{\coordindex{\tilde{u}^0}} K)\bigr]^2/16$.  \par The metric is Hermitian (but not necessarily K\"ahler) for the complex structure $I$ with $\mp i$-eigenspaces that are the span of the ${\coordindex{u^A}}$ and ${\coordindex{\tilde{u}^A}}$ coordinate vector fields respectively.  In fact,  the coordinates may be adapted to any locally integrable compatible complex structure $I$ such that the associated $2$-form $\Omega_{I} := g(I\cdot,\cdot)$ is \textit{self-dual}.  This choice corresponds to a pair of transverse self-dual foliations,  or equivalently,  a pair of non-intersecting surfaces in twistor space,  transverse to twistor lines.  If one desires that the complex structure preserves a real slice defined by an anti-holomorphic involution on twistor space,  then the two surfaces are required to be exchanged by the involution.  In that case one determines the other.  \par 
For Riemannian reality conditions,  it is sufficient that $K$ is real on the slice where $\tilde{u}^A = \bar{u}^{A}$.  This slice then carries a real-analytic Riemannian ASD Einstein metric and a compatible complex structure which is the restriction of $I$.  Taking $K$ to be real on the slice on which $(u^{A}, \tilde{u}^A)$ are real gives split-signature reality conditions.  $I$ of course does not preserve the real tangent bundle when restricted to this slice but the induced split-signature real analytic metric is \textit{para-Hermitian} for the real endomorphism $iI$.   \par 
The crucial observation that allows us to put the metric in the form (\ref{general_integrable}) is that we may rewrite Przanowski's equation as the following compatibility condition:
\begin{align}\label{compat}
\partial_{\coordindex{\tilde{u}^0}}\left(
\frac{\partial_{\coordindex{u^1}}\partial_{\coordindex{\tilde{u}^1}} K-\frac{2}{\Lambda}e^{-\Lambda K}}
{(\partial_{\coordindex{u^0}} K)^{2}}\right)
=\partial_{\coordindex{\tilde{u}^1}}\left(
\frac{\partial_{\coordindex{u^1}}\partial_{\coordindex{\tilde{u}^0}} K}{(\partial_{\coordindex{u^0}} K)^{2}}\right).  
\end{align}
\begingroup\setlength{\emergencystretch}{1em}
The geometric interpretation of this condition is as the $\tilde{\partial}$-closure of the $2$-form $(\partial_{\coordindex{u^0}} K)^{-2}\Omega_I$.  In fact,  given an ASD conformal structure and a choice of compatible complex structure,  such a scale always exists,  a fact also due to Przanowski \cite{Przan85}.  \par\endgroup 
Now work in coordinates $(z^{A},\theta^{A})$ where 
\begin{align}
z^{A} = u^{A},  \quad \theta^1 = \frac{2}{\Lambda^2 (\partial_{\coordindex{u^0}} K)} - \frac{u^0}{\Lambda}
\end{align}
and $\theta^0$ is defined (up to the addition of a function of the $u^{A}$) by 
\begin{align}\label{t0_def}
\partial_{\coordindex{\tilde{u}^0}} \theta^0 = \frac{2\partial_{\coordindex{u^1}}\partial_{\coordindex{\tilde{u}^0}} K}{\Lambda^2 (\partial_{\coordindex{u^0}} K)^{2}},  \quad \partial_{\coordindex{\tilde{u}^1}} \theta^0 = \frac{2(\partial_{\coordindex{u^1}}\partial_{\coordindex{\tilde{u}^1}} K-\frac{2}{\Lambda}e^{-\Lambda K})}{\Lambda^2 (\partial_{\coordindex{u^0}} K)^{2}}. 
\end{align}
The associated Jacobian is non-singular whenever $\partial_{\coordindex{\tilde{u}^0}} K\ne 0$ and hence wherever the metric is non-degenerate.  
\par In these coordinates $\phi = \Lambda \theta^{1} + z^0 = 2\Lambda^{-1}(\partial_{\coordindex{u}^0}K)^{-1}$ and the  metric is precisely
\begin{align}\label{pleb_rob} 
g = \frac{2}{\phi^2}(d \theta_{A} dz^{A} -(\partial_{\coordindex{u^B}} \theta_{A}) dz^{A} dz^{B}).  
\end{align}
The coordinate system is adapted to the SD foliation given by the $+i$-eigenspace of $I$.  \par 
Once we have put the metric into the above form\footnote{The metric is in what is often called Pleba\'nski-Robinson form.},  we may defer to,  for example,  \cite{DBook} and use that anti-self-duality implies the existence of functions $W^{A}(z^{B},\theta^{B})$ such that 
\begin{align}\label{component_equality}
\partial_{\coordindex{\theta^{(A}}} W_{B)}
=-\partial_{\coordindex{u^{(A}}} \theta_{B)}.
\end{align}
Finally,  the arguments of \cite{DM26} imply that there exists (possibly after redefining $W^{A}$) a function $W$ satisfying (\ref{W_potential}) and the LN equation $\Psi[W] = 0$.  
\begin{rem} 
The LN coordinates $(z^{A},\theta^{A})$ complete $u^{A} = z^{A}$ to a system of ``partial Darboux coordinates" for $(\partial_{\coordindex{u^0}} K)^{-2}\Omega_I$ in the sense $(\partial_{\coordindex{u^0}} K)^{-2}\Omega_I \propto dz_{A} \wedge d\theta^{A} + H_{AB} dz^{A} \wedge dz^{B}$ for some functions $H_{AB} = -H_{BA}$.  The coordinate transformation is therefore analogous to how coordinates adapted to Pleba\'nski's second heavenly equation may be obtained by completing the complex coordinates adapted to the first,  to a system of Darboux coordinates for the corresponding K\"ahler form as in \cite{Pl76}.  
\end{rem} 
In the following examples,  we can both integrate (\ref{t0_def}) and obtain $W$ explicitly: 
\begin{exmp}[Fubini-Study metric]\label{FS_exam}
We consider the Przanowski potential \cite{Hoeg} with $\Lambda = 1$ corresponding to the Fubini-Study metric on $\mathbb{CP}^2$ and analytically continue.  The Przanowski potential is given by:
\begin{align}
K = {\log[(1+|u|^2)(1+u^1\tilde{u}^1)]},  
\end{align}
where $|u|^2 = u^0\tilde{u}^0 +  u^1\tilde{u}^1$.  Perhaps confusingly,  the complex structure $I$ is not the one for which the metric is K\"ahler.  Rather,  the metric is K\"ahler with an anti-self-dual K\"ahler form.   \par Anyway,  we may integrate \textup{(\ref{t0_def})} explicitly.  One choice of integration constant gives:
\begin{align}
\theta^0 = \frac{2u^0\tilde{u}^1}{ \tilde{u}^0} + \frac{2\tilde{u}^1(u^0)^2}{1+u^1\tilde{u}^1},  \quad \theta^1 = {u^0} + \frac{2(1+u^1\tilde{u}^1)}{\tilde{u}^0}. 
\end{align}
From the above it is straightforward to check that the metric is
\begin{align}
g = \frac{2}{\phi^2}(d \theta_{A} dz^{A} + dz^0dz^0 - \frac{4\theta^0}{\phi}dz^0 dz^1 + \frac{2(\theta^0)^2}{\phi^2} dz^1dz^1), 
\end{align}
which is in Lipstein-Nagy form \textup{(\ref{general_integrable})} with $\Lambda = 1$ and 
\begin{align}
W = \frac{(\theta^0)^2}{2 \phi }.  
\end{align}
The Burns metric $g_B$ is given by 
\begin{align}
g_B = \frac{\phi^2}{4(z^0)^2}g,  
\end{align}
which is a scalar-flat K\"ahler metric for the complex structure $I$.  
\end{exmp} 
A generalisation is the following:
\begin{exmp}[One-loop universal hypermultiplet]
In \cite{APV},  the authors consider the Przanowski form of a one-loop approximation to the metric on the \textit{universal hypermultiplet} moduli space in Type IIA string theory on a rigid Calabi-Yau.  They obtain:
\begin{align}
K=\frac1\Lambda\log\frac{r^2}{r+c}
\end{align}
for a parameter $c$,  with $r$ determined implicitly by 
\begin{align}
\quad
r+c\log(r+c)
=\frac12(u^1+\tilde u^1)^2-\frac12(u^0+\tilde u^0),  
\end{align}
so the metric does not appear to be expressible in terms of elementary functions of $(u^A,\tilde{u}^A)$ unless $c = 0$.  Indeed,  in that case,  it gives the Fubini-Study metric \textup{(}and its Bergman cousin,  if $\Lambda < 0$\textup{)}.  
In the Riemannian setting,  it was shown in \cite{CS22} that this family of metrics \textup{(}in different coordinates,  and including $c = 0$\textup{)} gives all the complete Einstein metrics on $\mathbb{R}^4$ invariant under the standard action of $SO(2) \ltimes H$ on the last three coordinates,  where $H$ is the Heisenberg group.  \par Work in new Przanowski coordinates defined by
\begin{align}
v^0=u^0-(u^1)^2,\qquad v^1=u^1,
\end{align}
and similarly for the tilded coordinates.  
We may integrate \textup{(\ref{t0_def})} to 
\begin{align}
z^A=v^A,\quad 
\theta^0=-\frac{8(r+2c)\tilde v^1}{\Lambda},\quad
\theta^1=-\frac{4r+v^0}{\Lambda}.
\end{align}
We have $\phi=\Lambda\theta^1+z^0$,  and the metric takes
the LN form \textup{(\ref{general_integrable})} with potential
\begin{align}
W=\frac{(\theta^0)^2}{2\Lambda(\phi-8c)},
\end{align}
which one may check satisfies $\Psi[W] = 0$. 
\end{exmp}
\section{Lipstein-Nagy to Przanowski}
In this section,  we go in the opposite direction,  starting from a complex analytic metric written in Lipstein-Nagy form and writing it in Przanowski form.   
There is a choice of complex structure to which to adapt the Przanowski ansatz,  as emphasised in \cite{APV}.  \par  Start with a coordinate system $(z^{A},\theta^{A})$ in which an ASD Einstein metric takes the form (\ref{general_integrable}).  There is an infinite-dimensional freedom in choosing a surface in twistor space transverse to twistor lines,  equivalently a function $\eta(z,\theta)$ on $X$ satisfying 
\begin{align}
[\mathbf{E}_{A1'} + \eta \mathbf{E}_{A0'},  \mathbf{E}_{B1'} + \eta \mathbf{E}_{B0'}] = 0, 
\end{align}
which implies the twistor distribution lies tangent to the graph $\lambda = -\eta(z,\theta)$ and hence defines a surface in $Z$.  
This,  together with the twistor surface given by $\lambda = \infty$,  specifies a complex structure $I$ up to sign: we simply declare the corresponding self-dual foliations in $X$ to be the $\pm i$-eigenspaces.  \par 
We may find $\tilde{u}^{A}$ that give coordinates on the twistor surface,  which we may also view as the leaf space of the foliation defined by the span of the $\mathbf{E}_{A1'} + \eta \mathbf{E}_{A0'}$.  At this point we in fact have,  in the coordinates $(u^{A} := z^{A},\tilde{u}^{A})$ for $X$:
\begin{align}\label{deriv_iden}
\partial_{\coordindex{u^A}}
= \mathbf{E}_{A1'} + \eta \mathbf{E}_{A0'}. 
\end{align}
To get the Przanowski equation we must be discerning about the choice of $\tilde{u}^{A}$.  We use the formulae of H\"ogner from which we see that we should take the coordinate $\tilde{u}^1$ so that the contact form $\tau$ is proportional to $d\tilde{u}^1$ along $\lambda = -\eta$.   
We are free to do this since $\tau$ restricts (up to scale) to a $1$-form along this surface by virtue of descending (up to scale) to the twistor space.  On the other hand $\tilde{u}^0$ may be chosen freely.  \par
Take\footnote{There is a slight subtlety here in that $K$ will be undefined at points at which $\tau$ vanishes on the twistor surface.  However,  $\tau$ must be non-vanishing on a dense subset due to the contact condition.}
\begin{align}
\tau|_{\lambda = -\eta} = -\frac{\phi^2}{\Lambda}e^{-\Lambda K} d\tilde{u}^1
\end{align}
for some function (suggestively called) $K$ on $X$.  Indeed,  at this point it turns out that in the coordinates $(u^{A},\tilde{u}^{A})$,  the metric is in Przanowski form with potential $K$,  although it takes some showing: \par 
Evaluate the relation (\ref{tau_descent}) along $\lambda = -\eta$.  This yields:
\begin{align}
\partial_{\coordindex{u^A}}
\left(-\frac{\phi^2}{\Lambda}e^{-\Lambda K}\right)d\tilde{u}^1
=
2\Lambda 
(\partial_{\coordindex{\theta^A}}\partial_{\coordindex{\theta^0}} W)\,
\left(-\frac{\phi^2}{\Lambda}e^{-\Lambda K}\right)d\tilde{u}^1.
\end{align}
Evaluate the derivative on the left-hand-side for $A = 0,1$.  To do this,  note that we may calculate $\partial_{\coordindex{u^A}} \theta^{B}$ and hence $\partial_{\coordindex{u^A}} \phi$ from (\ref{deriv_iden}).  The two equations are:
\begin{align}\label{two_equations}
\partial_{\coordindex{u^0}} K
=\frac{2}{\Lambda\phi},  \quad 
\partial_{\coordindex{u^1}} K
=
\frac{2}{\phi}
\left(\eta+\Lambda\partial_{\coordindex{\theta^0}} W\right).
\end{align}
Since $\phi = \Lambda \theta^{1} + u^0$,  the first equation gives 
\begin{align}\label{eq_I}
\frac{2}{\phi^2}(\partial_{\coordindex{\tilde{u}^B}} \theta^1)d\tilde{u}^{B} = -(\partial_{\coordindex{u^0}}\partial_{\coordindex{\tilde{u}^B}} K) \,  d\tilde{u}^{B}.
\end{align}
We need to evaluate the contact form along $\lambda = -\eta$ in terms of the LN data.  An easy way to do this is to note that it must be in the span of $\mathbf{F}^{A0'}-\eta \mathbf{F}^{A1'}$ since it must annihilate the fields (\ref{deriv_iden}) by dint of descending to the surface.  To obtain the coefficients we plug in $\mathbf{E}_{A0'}$, which gives 
\begin{align}
\tau|_{\lambda=-\eta}
= 
\left[
\delta_{A}^{\, 0}-\phi^2\partial_{\coordindex{\theta^A}}
\bigg(
\phi^{-1}\bigg({\eta+\Lambda \partial_{\coordindex{\theta^0}} W\bigg)}
\bigg)
\right]
\left(\mathbf{F}^{A0'}-\eta \mathbf{F}^{A1'}\right).
\end{align}
Matching the above with the right-hand-side of the second equation of (\ref{two_equations}) and using our expression for the contact form in terms of Przanowski data:
\begin{align}
-\frac{\phi^2}{\Lambda}e^{-\Lambda K} d\tilde u^1
= (\mathbf{F}^{00'}-\eta \mathbf{F}^{01'})
-\frac{\phi^2}{2}\,d(\partial_{\coordindex{u^1}} K)
+\frac{\phi^2}{2}(\partial_{\coordindex{u^A}}\partial_{\coordindex{u^1}} K)\,\mathbf{F}^{A1'}.
\end{align}
Now $\mathbf{F}^{A1'} = du^{A}$ and $\mathbf{F}^{A0'}  - \eta \mathbf{F}^{A1'} = (\partial_{\coordindex{\tilde{u}^B}} \theta^{A})d\tilde{u}^{B}$ so:
\begin{align}\label{eq_II}
\frac{2}{\phi^2}
(\partial_{\coordindex{\tilde{u}^B}} \theta^0)d\tilde u^B
=
(\partial_{\coordindex{u^1}}\partial_{\coordindex{\tilde{u}^0}} K)d\tilde u^0+
\left(\partial_{\coordindex{u^1}}\partial_{\coordindex{\tilde{u}^1}} K-\frac{2}{\Lambda}e^{-\Lambda K}\right)d\tilde u^1.
\end{align}
Meanwhile,  (\ref{deriv_iden}) implies
\begin{align}\label{eq_III}
\partial_{\coordindex{u^B}} \theta_A
=-\partial_{\coordindex{\theta^B}} W_A
-\eta\epsilon_{AB}.
\end{align}
Finally,  substituting equations (\ref{eq_I},  \ref{eq_II},  \ref{eq_III}) into (\ref{general_integrable}),  one finds that the metric is in Przanowski form.  In particular,  $K$ satisfies the Przanowski equation (\ref{Prz_eqn}).  
\section{ASD Einstein metrics with symmetry and the $SU(\infty)$-Toda equation}
In this section,  we recall the $SU(\infty)$-Toda formulation of the ASD Einstein condition and recast some metrics adapted to this equation into `heavenly' form. 
The following is due to Tod:
\begin{thm}[Tod normal form \cite{Tod97}] 
Given a complex analytic ASD Einstein metric with $\Lambda \ne 0$ and non-null Killing vector field $\mathbf{V}$,  there exist local coordinates $(t,w,x,y)$ such that 
\begin{align}\label{SU_metric}
g= \frac{P}{w^2}(e^v(dx^2 + dy^2) + dw^2) + \frac{1}{Pw^2}(dt + \theta)^2
\end{align}
for some functions $P(w,x,y)$,  $v(w,x,y)$ and a $1$-form $\theta(w,x,y)$ related by 
\begin{align}
-4\Lambda P &= 2 - w\,\partial_{\coordindex{w}} v,  \label{toda_relation} \\
d\theta &= (\partial_{\coordindex{x}} P) dy \wedge dw + (\partial_{\coordindex{y}} P) dw \wedge dx + \bigl(\partial_{\coordindex{w}}(Pe^v)\bigr)\,dx \wedge dy,  \label{toda_relation_II}
\end{align}
where $v$ satisfies the $SU(\infty)$-Toda equation:
\begin{align}\label{SU_equation}
\partial_{\coordindex{x}}^2 v + \partial_{\coordindex{y}}^2 v + \partial_{\coordindex{w}}^2(e^{v})  = 0.  
\end{align}
Furthermore,  $\mathbf{V} = \partial_{\coordindex{t}}$.  
\end{thm} 
Note the coordinate freedom to replace $t$ by $t + \chi(w,x,y)$,  modifying $\theta$ appropriately.  \par 
The following reverses the steps of \cite{Tod06} equations (24)--(26):
\begin{prop}
Possibly after redefining $t$,  the metric \textup{(\ref{SU_metric})} may be put into Przanowski form \textup{(\ref{Prz_metric})} with 
\begin{align}
u^1 = x + iy, \quad u^0 = s + it, \nonumber \\
\label{prz_coords} \tilde{u}^1 = x - iy,  \quad \tilde{u}^0 = s - it,
\end{align}
where $s(w,x,y)$ is a function \textup{(}for which we may solve\textup{)} satisfying $\partial_{\coordindex{w}} s = P$ and $\partial_{\coordindex{x}}^2 s + \partial_{\coordindex{y}}^2 s + \partial_{\coordindex{w}}(Pe^{v}) = 0$,  with Przanowski function taken to be
 \begin{align}\label{Prz_from_Toda}
K = \frac{\log(4w^2) - v}{\Lambda}.  
\end{align}
Specifically,  after solving for $s$,  we must redefine $t$ so that $\theta = (\partial_{\coordindex{y}} s) dx - (\partial_{\coordindex{x}} s) dy$. \par 
Furthermore,  the Killing vector field is $\mathbf{V} = i(\partial_{\coordindex{u^0}} - \partial_{\coordindex{\tilde{u}^0}})$.   
\end{prop}
\begin{proof}
Take $s$ so $\partial_{\coordindex{w}} s = P$.  The Toda equation gives $(\partial_{\coordindex{x}}^2 s + \partial_{\coordindex{y}}^2 s + \partial_{\coordindex{w}}(Pe^{v})) = f(x,y)$,  and so we may redefine $s$ by adding a function of $(x,y)$ to satisfy $\partial_{\coordindex{x}}^2 s + \partial_{\coordindex{y}}^2 s + \partial_{\coordindex{w}}(Pe^{v}) = 0$.  Now,   
\begin{align}
d(\partial_{\coordindex{y}} s\,dx-\partial_{\coordindex{x}} s\,dy)
&=
\partial_{\coordindex{x}} P\,dy\wedge dw
+
\partial_{\coordindex{y}} P\,dw\wedge dx
-
(\partial_{\coordindex{x}}^2 s+\partial_{\coordindex{y}}^2 s)\,dx\wedge dy
=d\theta.
\end{align}
In particular,  we may redefine $t$ so that $\theta = (\partial_{\coordindex{y}} s)\,dx-(\partial_{\coordindex{x}} s)\,dy$.  Now,  taking $K$ as in (\ref{Prz_from_Toda}),  and using (\ref{toda_relation}), taking partial derivatives with respect to the Toda coordinates gives $\partial_{\coordindex{w}} K = -4P/w$.  From (\ref{prz_coords}) and $\partial_{\coordindex{w}} s = P$ we have
\begin{align}
\partial_{\coordindex{u^0}}
&=\frac{1}{2P}\partial_{\coordindex{w}}-\frac{i}{2}\partial_{\coordindex{t}},\quad \partial_{\coordindex{\tilde{u}^0}}
=\frac{1}{2P}\partial_{\coordindex{w}}+\frac{i}{2}\partial_{\coordindex{t}},  \nonumber \\
\partial_{\coordindex{u^1}}
&=\frac12\bigg(
\partial_{\coordindex{x}}-i\,\partial_{\coordindex{y}}
-\frac{\partial_{\coordindex{x}} s-i\,\partial_{\coordindex{y}} s}{P}\partial_{\coordindex{w}}
\bigg), \quad 
\partial_{\coordindex{\tilde{u}^1}}
=\frac12\bigg(
\partial_{\coordindex{x}}+i\,\partial_{\coordindex{y}}
-\frac{\partial_{\coordindex{x}} s+i\,\partial_{\coordindex{y}} s}{P}\partial_{\coordindex{w}}
\bigg).
\end{align}
Then 
\begin{align}
\partial_{\coordindex{u^0}} K=\partial_{\coordindex{\tilde{u}^0}} K=-\frac{2}{w}, \quad
\partial_{\coordindex{u^0}}\partial_{\coordindex{\tilde{u}^0}} K=\frac{1}{Pw^2},  \quad \partial_{\coordindex{u^0}}\partial_{\coordindex{\tilde{u}^1}} K
=-\frac{\partial_{\coordindex{x}} s+i \partial_{\coordindex{y}} s}{Pw^2},
 \\ 
\partial_{\coordindex{u^1}}\partial_{\coordindex{\tilde{u}^0}} K
=-\frac{\partial_{\coordindex{x}} s-i \partial_{\coordindex{y}} s}{Pw^2}, \quad 
\partial_{\coordindex{u^1}}\partial_{\coordindex{\tilde{u}^1}} K
-\frac{2}{\Lambda}e^{-\Lambda K}
=
\frac{(\partial_{\coordindex{x}} s)^2+(\partial_{\coordindex{y}} s)^2}{Pw^2}
+
\frac{Pe^v}{w^2},
\end{align}
from which it is straightforward to check that the metrics (\ref{Prz_metric}) and (\ref{SU_metric}) are equal.  The Przanowski equation holds by virtue of the ASD Einstein condition.  
\end{proof}
If we now apply the transformation in \S \ref{PtoLN} to put the metric into LN form,  we have $z^0 = s + it,  \ z^1 = x + iy,  \ \theta^1 = -w/\Lambda^2 - (s+it)/\Lambda$.  Defining
\begin{align}\label{Toda to LN}
\partial_{\coordindex{w}} \theta^0
=-\frac{\partial_{\coordindex{x}} s-i \partial_{\coordindex{y}} s}{\Lambda^2},
\qquad
\partial_{\coordindex{x}} \theta^0
+i\,\partial_{\coordindex{y}} \theta^0
=\frac{Pe^v}{\Lambda^2},
\qquad
\partial_{\coordindex{t}} \theta^0=0,
\end{align}
the Killing vector becomes 
\begin{align}
\mathbf{V} = i\bigg(\partial_{\coordindex{z^0}} - \frac{1}{\Lambda} \partial_{\coordindex{\theta^1}}\bigg).  
\end{align}
 \begin{exmp}[ASD Taub-NUT with cosmological constant]\label{ATN}
The ASD Taub-NUT metric \cite{GP78} \textup{(}also known as the Pedersen metric \cite{Pedersen}\textup{)} with cosmological constant $\Lambda$ is,  using coordinates $(\rho,\vartheta,\varphi,\psi)$,   
\begin{align}\label{Pedersen_metric}
g={}&\frac{\rho^2-l^2}{\Delta}\,d\rho^2
+\frac{4l^2\Delta}{\rho^2-l^2}
 (d\psi+\cos\vartheta\,d\varphi)^2
+(\rho^2-l^2)(d\vartheta^2+\sin^2\vartheta\,d\varphi^2), \\ 
\Delta={}&(\rho-l)^2
 \bigl[1-2\Lambda(\rho-l)(\rho+3l)\bigr].
\end{align}
Here,  $l$ is a constant called the NUT parameter.  In the Riemannian setting,  if $\Lambda < 0$,  $l > 0$,  and on the region $l \le \rho < \infty$ with $(\vartheta,  \varphi, \psi)$ the usual Euler angles on $S^3$,  $g$ defines a complete metric on $\mathbb{R}^4$ with a removable coordinate singularity at the origin $\rho = l$.  \par 
Indeed,  the submanifolds of fixed radial coordinate $\rho$ are Berger spheres so $g$ has $SU(2) \times U(1)$ symmetry.  We shall put the metric into $SU(\infty)$-Toda form adapted to the Killing vector $\partial_{\coordindex{t}} = 2\Lambda \partial_{\coordindex{\psi}}$. 
This was earlier done by Nozawa \cite{N17} in a slightly different fashion.  Here we take coordinates $(w,x,y,t)$ given by \begin{align}
w=\frac{1}{4\Lambda l(\rho-l)},\quad
t=\frac{\psi-\varphi}{2\Lambda},\quad
x+iy=\cot\frac{\vartheta}{2}\,e^{i\varphi}.
\end{align}
Geometrically,  $(x,y)$ are stereographic coordinates on the base of the Hopf fibration of the Berger sphere,  $t$ is a fibre coordinate,  and $w$ is a new radial coordinate which,  assuming $ \Lambda < 0$,  maps the $l < \rho < \infty$ patch to $-\infty < w < 0$ with $\rho = l$ \textup{(}the nut\textup{)} at $-\infty$.     \par 
We check using the Maple {\texttt{DifferentialGeometry}} package that the metric becomes \textup{(\ref{SU_metric})} with 
\begin{align}
e^v &=\frac{8(w-1)^2-l^{-2}\Lambda^{-1}-8}{2(1+x^2+y^2)^2},\ \ 
P=\frac{8w+l^{-2}\Lambda^{-1}}{\Lambda(16(w-1)^2-16-2l^{-2}\Lambda^{-1})}, \nonumber \\
\theta & =\frac{xdy-ydx}{\Lambda(1+x^2+y^2)},  
\end{align}
and that the above satisfy \textup{(\ref{toda_relation}} - \textup{\ref{SU_equation})}.  \par 
Next,  we cast the metric into LN form.  We may take,  locally,  
\begin{align}
s=\frac1{4\Lambda}(
(1+\gamma)\log(w-1-\gamma)
+(1-\gamma)\log(w-1+\gamma)
-2\log(1+x^2+y^2)),
\end{align}
where $\gamma = (1+1/(8\Lambda l^2))^{1/2}$.  This satisfies $\partial_{\coordindex{w}} s = P$,  $\partial_{\coordindex{x}}^2 s + \partial_{\coordindex{y}}^2 s + \partial_{\coordindex{w}}(Pe^v) = 0$ and $t$ is already normalised so $\theta = (\partial_{\coordindex{y}} s) dx - (\partial_{\coordindex{x}} s) dy$.  We may then integrate \textup{(\ref{Toda to LN})} explicitly,  which gives:
\begin{align}
\theta^0=
\frac{(8w+l^{-2}\Lambda^{-1})(x-iy)}
{8\Lambda^3(1+x^2+y^2)}.  
\end{align}
Finally,  we now use $z^0 = s + it,  \ z^1 = x + iy,  \ \theta^1 = -w/\Lambda^2 - (s+it)/\Lambda$ and the relation
\begin{align}
ds = P dw - \frac{x dx + y dy}{\Lambda(1+x^2+y^2)}
\end{align}
to check that the metric \textup{(\ref{general_integrable})} with the surprisingly simple potential
\begin{align}
W=\frac{4\Lambda l^2(\theta^0)^2(1+\Lambda\phi)}
          {8\Lambda^2l^2\phi-1}
\end{align}
is \textup{(\ref{SU_metric})} with the $SU(\infty)$-Toda data above,  and hence is the analytically continued ASD Taub-NUT metric with cosmological constant $\Lambda$.  \par 
Assuming $\gamma \ne 0$,  so the metric is not conformally flat,  there is a curvature singularity along the complex hypersurface $8\Lambda^2 l^2 \phi = 1$.  The equations defining the Riemannian real slice in terms of our holomorphic coordinates somewhat obscure the geometry but note that along this slice $ -\Lambda\phi = w$ is real \textup{(}assuming real $\Lambda$\textup{)}.  We do,  however,   obtain a real analytic metric of signature $(2,2)$ on the slice with $(z^{A},\theta^{A})$ real.  \par 
After taking the $l \to \infty$ limit of this family of potentials,  one obtains the Fubini-Study metric of Example \textup{\ref{FS_exam}} with $\Lambda = 1$,   if one works in LN coordinates
\begin{align}
\tilde{z}^0 &= 1-\exp(-z^0),  \quad \tilde{z}^1 = z^1,  \\ \tilde{\theta}^0 &= \exp(-2z^0)\theta^0, \quad  \   \tilde{\theta^1} = \exp(-z^0)\phi - 1 + \exp(-z^0). 
\end{align}
That the Fubini-Study metric can be obtained as the $l \to \infty$ limit of \textup{(\ref{Pedersen_metric})} was observed in \cite{GP78}.    \par 
 \end{exmp}
 
\section{Linearised solutions and contour integral formulae}
Let us compute the linearised Lipstein-Nagy equation.  Let $W$ be a solution to the non-linear equation and write
\begin{align}
W_\epsilon = W + \epsilon h
\end{align}
so $h$ is a first-order perturbation.  Then
\begin{align}\label{LLN}
\frac{d}{d\epsilon}\bigg|_{\epsilon=0}\hspace{-1mm}\Psi[W_\epsilon] = \mathcal{Q}(h) + 2\Lambda \epsilon^{AB} (\partial_{\coordindex{\theta^0}}\partial_{\coordindex{\theta^A}} W) (\partial_{\coordindex{\theta^B}} h) = 0
\end{align}
is the linearised Lipstein-Nagy equation with respect to the background defined by $W$.  Substituting the perturbed potential into the metric (\ref{general_integrable}) gives 
\begin{align}
   g_\epsilon = g + \frac{2\epsilon}{\phi^2}
   \partial_{\coordindex{\theta^B}}
   \bigg(\phi \partial_{\coordindex{\theta^A}} h-3\Lambda \delta_{A}^{\, 1} h\bigg)\,dz^A dz^B, 
\end{align}
which satisfies the ASD Einstein equations to first order in $\epsilon$.  \par 
In \cite{DM26} it was explained how to obtain series expansions for twistor functions in the parameter $\lambda$.  It is the natural analogue of Dunajski-Mason recursion \cite{DM00} in our setup.  Suppose $f$ is holomorphic on $X \times \{r_0 < |\lambda| < r_1\}$
for some positive radii $r_0 < r_1$ and descends to a twistor function on an open subset of $Z$. 
We may Laurent expand $f = \sum_{n \in \mathbb{Z}} \lambda^{-n}f_n$,  where the $f_n$ are functions on $X$.  The twistor condition $\mathbf{L}_A(f)=0$ defines a recursion relation between the Laurent coefficients  defined in terms of the vector fields (\ref{Evectors}):
\begin{align}
\label{recursionrel}
\mathbf{E}_{A0'}(f_{n+1})=\mathbf{E}_{A1'}(f_n)+(n-1)(\partial_{\coordindex{\theta^A}} \beta)f_{n-1}.
\end{align}
We may generalise this procedure to obtain a recursion relation for sections of line bundles $\mathcal{O}(k)$,  $k \in \mathbb{Z}$,  on twistor space.  The descent condition (\ref{ok_section_practical}) yields a recursion relation for the coefficients of the expansion $\psi = \sum_{n \in \mathbb{Z}} \lambda^{-n}\psi_n$ representing a section of $\mathcal{O}(k)$:
\begin{align}
\mathbf{E}_{A0'}(\psi_{n+1})
=\mathbf{E}_{A1'}(\psi_n)
-k\Lambda 
(\partial_{\coordindex{\theta^A}}\partial_{\coordindex{\theta^0}} W) \psi_n +(n-1)(\partial_{\coordindex{\theta^A}} \beta)\psi_{n-1}.
\end{align}
Recall the discussion in \S \ref{twistors}:  $\psi$ representing a section of $\mathcal{O}(-1)$ restricted to an $\alpha$-surface $\Sigma$ gave the value of a parallel section of $\nabla^{\Sigma}$ with respect to the trivialisation $\sigma = e_{1'} - \lambda e_{0'}$ for $L_\Sigma$.  Now,  this is not a holomorphic trivialisation in the neighbourhood of $\lambda = \infty$ but $e_{1'}/\lambda - e_{0'}$ is.  So $\psi$ defines a holomorphic section over $\infty$ if and only if the $n < 1$ terms in the Laurent expansion vanish.  It follows $\psi$ representing a section of $\mathcal{O}(k)$ extending holomorphically over $\infty$ takes the form  $\psi = \sum_{n \ge -k} \lambda^{-n}\psi_n$.  \par 
A $\psi$ representing a section of $\mathcal{O}(2)$ defined on  $X \times \{r_0 < |\lambda| < r_1\}$ defines a class in $\check{H}^1(Z,\mathcal{O}(2))$.  Working locally in spacetime,  the corresponding Hamiltonian vector field (see \S \ref{w1}) defines an infinitesimal deformation of the holomorphic structure on twistor space preserving the contact structure.  On the spacetime side,  such data corresponds to a linearised ASD Einstein deformation  as follows:
\begin{prop}Let 
$\psi = \sum_{n\in \mathbb{Z}} \lambda^{-n}\psi_n$ be a holomorphic function on $X \times \{r_0 < |\lambda| < r_1\}$ representing a section of $\mathcal{O}(2)$ on $Z$.  Then,  $\psi_1$ satisfies the linearised LN equation.  
\end{prop}
\begin{proof}
For $n = 1$ we have 
\begin{align}
\mathbf{E}_{A0'}(\psi_{2})
=\mathbf{E}_{A1'}(\psi_1)
-k\Lambda 
(\partial_{\coordindex{\theta^A}}\partial_{\coordindex{\theta^0}} W) \psi_1.  
\end{align}
Now differentiate both sides by $\mathbf{E}_{B0'}$ and skew-symmetrise.  Using the commutation relation $[\mathbf{E}_{B0'},  \mathbf{E}_{A1'}] = [\mathbf{E}_{A0'},  \mathbf{E}_{B1'}]$ gives the linearised LN equation for $k = 2$.  
\end{proof}
By Cauchy's theorem we can of course extract the $n = 1$ coefficient by contour integral formula:
\begin{align}
\psi_1 = \frac{1}{2\pi i}\oint_{\Gamma} \psi \ d\lambda,
\end{align}
where $\Gamma$ is a positively oriented contour winding once around the annulus $\{r_0 < |\lambda| < r_1\}$.  \par 
Conversely,  the difference between two sections (constructed via the recursion procedure,  for example) with the same $n = 1$ term is trivial in $\check{H}^1(Z,\mathcal{O}(2))$.  To see this,  the recursion relation implies that the $n < 1$ and $n > 1$ parts of the difference,   which extend over $\lambda = 0$ and $\lambda = \infty$ respectively,  separately descend to $Z$.  The difference is therefore a coboundary.   \par However,  non-zero solutions to the linearised Lipstein-Nagy equation can of course give rise to trivial classes in $\check{H}^1(Z,\mathcal{O}(2))$.    Indeed,  if we take $\psi_0 = z^{A}$ and set $\psi_n = 0$ for $n < 0$,  then we get a solution to (\ref{LLN}):
\begin{align}
\psi_1
=\theta^A-2\Lambda z^A \partial_{\coordindex{\theta^0}} W .  
\end{align}
However,  any corresponding $\psi = \sum_{n \ge 0}^{\infty} \lambda^{-n}\psi_n$ is holomorphic over $\lambda = \infty$ and therefore represents a trivial class in $\check{H}^1(Z,\mathcal{O}(2))$.  This is to be expected: one may check the linearised solution above is ``pure gauge" in that it corresponds to the infinitesimal diffeomorphism generated by the vector field $-2\Lambda z^A \partial_{\coordindex{\theta^0}}$.    \par 
From sections of $\mathcal{O}(k)$ we may similarly obtain formulae for solutions of
\begin{align}\label{family_PDE}
\mathcal{Q}(h) + k\Lambda \epsilon^{AB} (\partial_{\coordindex{\theta^0}}\partial_{\coordindex{\theta^A}} W) (\partial_{\coordindex{\theta^B}} h) = 0
\end{align}
for $k \in \mathbb{Z}$.
In fact,  this is the conformal wave equation for $\phi h$ when $k = -2$.  \par 
It is worth discussing the related work of H\"ogner \cite{Hoeg} on the Przanowski equation,  and Araneda \cite{A25}.  In their setting,  the chosen spin frame defines an integrable complex structure.  For representatives of sections of $\mathcal{O}(k)$,  they obtain a one-step recursion relation for the coefficients of the analogous expansions.  Furthermore,  each term in the expansion is shown to satisfy a second-order linear homogeneous scalar PDE.  Here we have only obtained a PDE satisfied by $\psi_1$,  and the analogous PDEs satisfied by the other Laurent coefficients are coupled.  It is not clear to the author how to get around this limitation without picking a complex structure and an associated spin frame (therefore changing the parameter with respect to which the expansions are taken). \par  
If one uses the Przanowski or Lipstein-Nagy normal forms to view the metric as a field on a fixed background distinguished by the coordinate system,  then the metric perturbation associated to a solution of the linearised Przanowski equation remains Hermitian for a fixed (the distinguished) complex structure,  whereas a solution to the linearised Lipstein-Nagy equation need not preserve Hermiticity for such a choice.  The perturbation instead preserves the property that the distinguished foliation is SD. 
\section{The algebra of twistor Hamiltonians}\label{w1}
In the Ricci-flat case the twistor space of an ASD Einstein metric carries an $\mathcal{O}(2)$-valued $2$-form (dual to what is called the infinity twistor) relative to a fibration $Z \to \mathbb{CP}^1$.  This defines a degenerate Poisson bracket on sections of $\mathcal{O}(2)$.  Recently \cite{AMS22},  it was observed that what was previously being studied under the name $Lw_{1+\infty}$ or celestial symmetry algebra could be interpreted as arising via this bracket. \par 
The contact structure on $Z$ associated to an ASD Einstein metric with $\Lambda \ne 0$ also allows us to construct an algebra of sections of $\mathcal{O}(2)$.   We briefly review the notion of a \textit{Jacobi bracket} associated with our contact form $\tau$: \par 
Each local section $\psi$ of $\mathcal{O}(2)$ determines a unique vector field $\mathbf{V}_\psi$ determined by:
\begin{align}
\tau(\mathbf{V}_\psi) = \psi,  \quad [\mathbf{V}_\psi, H] \subseteq H,
\end{align}
where $H = \ker \tau$ is the contact distribution.  \par 
By construction $\mathbf{V}_\psi$ is a contact vector field;  it preserves the contact distribution.  \par Conversely,  every contact vector field arises locally this way.
We define the Jacobi bracket $\{\psi,\chi\}$ by
\begin{align}
\mathbf{V}_{\{\psi,\chi\}}:= [\mathbf{V}_\psi,\mathbf{V}_\chi]
\end{align}
so
\begin{align}
\{\psi, \chi\} = \tau([\mathbf{V}_\psi,\mathbf{V}_\chi]).
\end{align}
We now determine this bracket explicitly on representative functions $\psi,\chi$ satisfying (\ref{ok_section_practical}) for $k = 2$ (this is of course a slight abuse of notation as we are identifying sections of $\mathcal{O}(2) \to Z$ as functions on $X \times \mathbb{CP}^1$).  \par
The first step is a formula for the contact vector field $\mathbf{V}_\psi$ associated to a Hamiltonian $\psi$.  
We have
\begin{align}\label{dt}
d\tau = 
2\Lambda^2(\partial_{\coordindex{\theta^0}}^2 W)\,
d\theta^0\wedge d\theta^1
+2 \Lambda d\theta^1\wedge d\lambda + ...
\end{align}
where we have omitted terms with a $dz^{A}$ factor.  Now $\mathbf{V}_\psi$ Lie deriving the contact distribution requires
\begin{align}\label{descend}
\mathcal{L}_{{\mathbf{V}_{\psi}}}\tau = \iota_{{\mathbf{V}_\psi}}d\tau + d\psi = \eta \tau
\end{align}
for some function $\eta$.  It is straightforward to solve the above and obtain
\begin{align}
\eta = \partial_{\coordindex{\theta^0}} \psi + \Lambda (\partial_{\coordindex{\theta^0}}^2 W)\partial_{\coordindex{\lambda}} \psi
\end{align}
by evaluation of (\ref{descend}) on $\partial_{\coordindex{\lambda}}$ and $\partial_{\coordindex{\theta^0}}$.  Representative vector fields on $X \times \mathbb{CP}^1$ can be chosen to satisfy $\mathbf{V}_\psi(z^{A}) = 0$ since we have the freedom to add vector fields in the twistor distribution.  Solving (\ref{descend}) and $\tau(\mathbf{V}_\psi) = \psi$ for $\mathbf{V}_\psi$ (using (\ref{dt}))  gives
\begin{align}
2\Lambda 
\mathbf{V}_\psi
={}&\phi\left(
\partial_{\coordindex{\theta^0}}\psi\,\partial_{\coordindex{\theta^1}}
-\partial_{\coordindex{\theta^1}}\psi\,\partial_{\coordindex{\theta^0}}
\right)
+\Lambda\left(
\lambda-\Lambda\partial_{\coordindex{\theta^0}}W
+\phi\,\partial_{\coordindex{\theta^0}}\partial_{\coordindex{\theta^1}}W
\right)\hspace{-1.5mm}
\left(
\partial_{\coordindex{\theta^0}}\psi\,\partial_{\coordindex{\lambda}}
-\partial_{\coordindex{\lambda}}\psi\,\partial_{\coordindex{\theta^0}}
\right) \nonumber \\
&+\left(1-\Lambda\phi\,\partial_{\coordindex{\theta^0}}^{\,2}W\right)\hspace{-1.5mm}
\left(
\partial_{\coordindex{\theta^1}}\psi\,\partial_{\coordindex{\lambda}}
-\partial_{\coordindex{\lambda}}\psi\,\partial_{\coordindex{\theta^1}}
\right)
+2\Lambda\psi\left(
\partial_{\coordindex{\theta^0}}
+\Lambda\partial_{\coordindex{\theta^0}}^{\,2}W\,\partial_{\coordindex{\lambda}}
\right).
\end{align}
A tedious computation calculating the commutator and evaluating against $\tau$ then gives
\begin{align}
2\Lambda\{\psi,\chi\}
={}&\phi\left(
\partial_{\coordindex{\theta^0}}\psi\,\partial_{\coordindex{\theta^1}}\chi
-\partial_{\coordindex{\theta^1}}\psi\,\partial_{\coordindex{\theta^0}}\chi
\right) +\left(1-\Lambda\phi\,\partial_{\coordindex{\theta^0}}^{\,2}W\right)
\left(
\partial_{\coordindex{\theta^1}}\psi\,\partial_{\coordindex{\lambda}}\chi
-\partial_{\coordindex{\lambda}}\psi\,\partial_{\coordindex{\theta^1}}\chi
\right) \nonumber \\ \nonumber
&+\Lambda\left(
\lambda-\Lambda\partial_{\coordindex{\theta^0}}W
+\phi\,\partial_{\coordindex{\theta^0}}\partial_{\coordindex{\theta^1}}W
\right)
\left(
\partial_{\coordindex{\theta^0}}\psi\,\partial_{\coordindex{\lambda}}\chi
-\partial_{\coordindex{\lambda}}\psi\,\partial_{\coordindex{\theta^0}}\chi
\right)\\ 
&+2\Lambda\left(
\psi\,\partial_{\coordindex{\theta^0}}\chi
-\chi\,\partial_{\coordindex{\theta^0}}\psi
\right)
+2\Lambda^2\partial_{\coordindex{\theta^0}}^{\,2}W
\left(
\psi\,\partial_{\coordindex{\lambda}}\chi
-\chi\,\partial_{\coordindex{\lambda}}\psi
\right).
\end{align}
The algebra that might be called $\Lambda$-deformed $Lw_{1+\infty}$ is,  on a conformally flat background,  a subalgebra of contact Hamiltonians on twistor space generated by Hamiltonians $w_{m,a}^{p}$ (where $p \pm m-1\in\mathbb Z_{\ge0}$ and $a\in\mathbb Z+p$) satisfying:
\begin{align}\label{taylor_zhu}
2\Lambda\{w^p_{m,a},w^q_{n,b}\}
={}&\bigl(m(q-1)-n(p-1)\bigr)
w^{p+q-2}_{m+n,a+b} \nonumber \\
&-\Lambda\bigl(a(q-2)-b(p-2)\bigr)
w^{p+q-1}_{m+n,a+b}.
\end{align}
See \cite{BBHKMS24} for full details.  
\section{ASD Taub-NUT in a curved background and explicit twistor Hamiltonians}\label{ASDTaub_explicit}
In this section,  we revisit Example \ref{ATN} where we found the solution of the LN equation corresponding to the ASD Taub-NUT metric with cosmological constant.  In particular,  we will find functions $Z^0,Z^1,\zeta$ on $\mathbb{P}S' \cong X \times \mathbb{CP}^1$ descending to local coordinates on the twistor space $Z$.  Remarkably,  these functions are elementary in the coordinates $(z^{A},\theta^{A},\lambda)$.  This gives a $\Lambda \ne 0$ parallel of some of the results of \cite{Kim26}.  Next,  we explain how these functions can be used to write down contact Hamiltonians on twistor space,  and demonstrate in an example that we can use contour integration to obtain explicit solutions of the linearised LN equation on this background.   The twistor theory of this metric has already been developed in detail \cite{Pedersen,Z02,BH26}.  The results of this section should be thought of as complementary,  and a demonstration of how the machinery we have developed in this article may be put into practice in a setting of interest. \par 
We employed the \texttt{DifferentialGeometry} package and \texttt{pdsolve} function in Maple to explicitly integrate the twistor distribution.  When performing the integration it is best to use coordinates $(z^0,z^1,\theta^0,\phi)$.  Then,  the metric becomes invariant under translations in the $z^{A}$ coordinates.  It follows that the corresponding coordinate vector fields (in these coordinates, thought of as vector fields on $X \times \mathbb{CP}^1$) descend to twistor space.  In particular,  when searching for twistor functions $Z^{A}$,  we may employ the following ansatz: 
\begin{align}
Z^{A} = z^{A} + F^{A}(\phi,\theta^0,\lambda).  
\end{align}
For the remainder of the section we set $\Lambda = -1$ and so $\gamma^2 = 1 - 1/(8l^2)$.
One choice of integration constants gives the following functions annihilated by the vector fields $\mathbf{L}_{A}$:
\begin{align}
Z^0 &= z^0-\frac12\sum_{\pm}(1\pm\gamma)\log \frac{(\phi-1+\gamma^2)\lambda+(\phi-1\mp\gamma)\theta^0}{(\phi-1+\gamma^2)\lambda+2(\phi-1)\theta^0},\\
Z^1 &= z^1-\frac{(\phi-1+\gamma^2)[\phi(\phi-1+\gamma^2)\lambda+(\phi^2-1+\gamma^2)\theta^0]}{[(\phi-1+\gamma^2)\lambda+(\phi-1)\theta^0]^2-\gamma^2(\theta^0)^2},\\
\zeta &= \frac{(\phi-1+\gamma^2)[(\phi-1+\gamma^2)\lambda+2(\phi-1)\theta^0]}{[(\phi-1+\gamma^2)\lambda+(\phi-1)\theta^0]^2-\gamma^2(\theta^0)^2}.
\end{align}
Twistor functions are then given by functions $f(Z^0,Z^1,\zeta)$.  Fixing a spacetime point and varying $\lambda$ to sweep out a twistor line,  $(Z^0,Z^1,\zeta) = (z^0,z^1,0)$ at $\lambda = \infty$.  For a generic spacetime point,  the functions $\zeta,  Z^1$ each have a pair of poles $\lambda_{\pm}$ in $\lambda$ whereas $Z^0$ is not single-valued on the twistor line and branches at three points:
\begin{align}
\lambda_\pm=-\frac{(\phi-1\pm\gamma)\theta^0}{\phi-1+\gamma^2},
\qquad
\lambda_0=-\frac{2(\phi-1)\theta^0}{\phi-1+\gamma^2}.
\end{align}
\begin{rem} The approach in \cite{BH26} is somewhat different: the curved metric is viewed as corresponding to a deformation of the complex structure of flat twistor space.  The twistor coordinates are expressed in terms of non-holomorphic functions in the complex coordinates on flat twistor space \textup{(}which are annihilated by their deformed Dolbeault operator,  hence holomorphic with respect to the deformed complex structure\textup{)}.  \end{rem}
In order to find functions on $X \times \mathbb{CP}^1$ representing sections of $\mathcal{O}(2)$,  and hence contact Hamiltonians,  we choose a trivialisation of $\mathcal{O}(2)$ over an open subset of $Z$.  In practice we must find a non-vanishing function $\rho$ satisfying (\ref{ok_section_practical}) for $k = 2$ and then use it as a prefactor.  That is,  given such a $\rho$,  a
$\psi$ satisfying (\ref{ok_section_practical}) with $k = 2$ may be written
\begin{align}
\psi = \rho f(Z^0,Z^1,\zeta)
\end{align}
for a function $f$.  One such choice of $\rho$ is
\begin{align}\label{easy_rho}
\rho = \lambda+
\frac{2(\phi-1)\theta^0}{\phi-1+\gamma^2}.
\end{align}
As a section of $\mathcal{O}(2)$ it has a simple zero at $\lambda = \infty$ and $\lambda = \lambda_0$ on each twistor line.  
Via contour integration,  we now have a solution generating method for the linearised LN equation on this background.  A particularly simple example is obtained from 
\begin{align}
\psi = \frac{\rho}{\zeta^2}.  
\end{align}
We evaluate
\begin{align}
h := \frac{1}{2\pi i}\oint_{\Gamma}
\psi \ d\lambda = \frac{1}{2\pi i}
\oint_{\Gamma}
\frac{
(
\left[(\phi-1+\gamma^{2})\lambda+(\phi-1)\theta^{0}\right]^{2}
-\gamma^{2}(\theta^{0})^{2}
)^{2}
}{
(\phi-1+\gamma^{2})^{3}
\left[(\phi-1+\gamma^{2})\lambda+2(\phi-1)\theta^{0}\right] 
}d\lambda.  
\end{align}
Here we take the contour $\Gamma$ to contain the finite pole at 
\begin{align}
\lambda=\frac{2(1-\phi)\theta^0}{\phi-1+\gamma^2}
\end{align}
and we obtain
\begin{align}
h
=\frac{(\theta^{0})^{4}
((\phi-1)^2-\gamma^{2})^2}
{(\phi-1+\gamma^{2})^{4}},
\end{align}
which one may readily check satisfies the linearised LN equation on this background.  Since $\psi$ is independent of $Z^0,Z^1$,  the perturbation respects translations in $z^{A}$ in the coordinates $(z^A, \theta^0,\phi)$.  It is,  however,  a genuine ASD Einstein perturbation: Writing $I_\epsilon,  J_\epsilon$ for the usual \cite{PR86} quadratic and cubic scalars of the Weyl spinor of the perturbed metric $g_\epsilon$,  using Maple we check the Type D property is not preserved:
\begin{align}
\frac{d^2}{d\epsilon^2} \bigg|_{\epsilon = 0}(I_\epsilon^3 - 27J_\epsilon^2) \ne 0
\end{align}
at the point $z^0 = z^1 = \theta^0 = 1,  \,  \theta^1 = 0$.
The coordinates $(Z^0,Z^1,1/(2\zeta))$ are in fact Darboux for the contact structure in an appropriate trivialisation.  A representative contact $1$-form on $Z$ is given by:
\begin{align}
{\tau}/(2\rho)=dZ^0+ \frac{1}{2\zeta}dZ^1,
\end{align}
with $\rho$ as in (\ref{easy_rho}).  In particular,  the Jacobi bracket of sections of $\mathcal{O}(2)$ given by $\psi = \rho f(Z^0,Z^1,\zeta)$ and $\chi = \rho \hat{f}(Z^0,Z^1,\zeta)$ takes the canonical form:
\begin{align}
\{ \psi,  \chi \} 
={} \rho \bigg[ &\frac{\zeta}{2}
\bigl((\partial_{\coordindex{\zeta}} f) (\partial_{\coordindex{Z^0}} \hat{f})-(\partial_{\coordindex{Z^0}} f)(\partial_{\coordindex{\zeta}} \hat{f})\bigr) \nonumber \\
&+\zeta^2
\bigl((\partial_{\coordindex{Z^1}} f)(\partial_{\coordindex{\zeta}} \hat{f})-(\partial_{\coordindex{\zeta}} f) (\partial_{\coordindex{Z^1}} \hat{f})\bigr) 
+\frac12\left(f\,\partial_{\coordindex{Z^0}} \hat{f}-\hat{f}\,\partial_{\coordindex{Z^0}} f\right) \bigg].
\end{align}
For this metric,  a subalgebra generated by physically relevant modes reducing to (\ref{taylor_zhu}) in the conformally flat limit ($\gamma = 0$ in our notation) was identified in \cite{BH26},  and it remains to interpret them in terms of our coordinates.  
\section{Conclusion}
We have continued our study,  beginning in \cite{DM26},  of the equation of Lipstein and Nagy.  
We hope that many of the results obtained here can be applied in the study of gravitational scattering on ASD backgrounds,  or more generally,  the twistor theory and deformation theory of quaternion-K\"ahler metrics in four dimensions.  The ASD Taub-NUT example shows that non-linear twistor theory can be made surprisingly computationally tractable once the metric is cast into LN form.  
The transformations we have provided between the Tod,  Przanowski and LN forms of an ASD Einstein metric should be applicable beyond the examples considered here.   \par 
A particularly concrete next step would be to adapt Miller's perturbiner methods \cite{Miller1,Miller2} to study the LN equation.  \par 
Another interesting point to resolve is the relation between the family of PDEs (\ref{family_PDE}) and the family of Teukolsky equations in \cite{A25}.  Indeed,  (\cite{A25},  2.18) with $w = 1,  p = 0$ was shown there to be equivalent to the linearised Przanowski equation studied in \cite{Hoeg}.  Since a solution to the equation (\ref{LLN}) also represents (locally) the general linearised ASD Einstein perturbation,  a similar statement will be able to be made for the linearised LN equation,  although the precise details will be left to future work.  \par
Finally,  in the Ricci-flat case,  the second heavenly equation generalises to a system of $n(2n-1)$ PDEs for a single scalar potential.   It would be interesting to determine whether there is an analogous generalisation of the Lipstein-Nagy equation governing quaternion-K\"ahler metrics in $4n$ dimensions.  


\begin{thebibliography}{99}\raggedright

\bibitem{AAS26}
T. Adamo, B. Araneda, S. Seet,
{\em The dual twistor theory of self-dual black holes},
arXiv:2601.05037, (2026).

\bibitem{AASS26}
T. Adamo, B. Araneda, S. Seet, A. Sharma,
{\em Schwarzschild black holes from twistor space},
arXiv:2607.06236, (2026).

\bibitem{AMS22}
T. Adamo, L. J. Mason, A. Sharma,
{\em Celestial $w_{1+\infty}$ Symmetries from Twistor Space},
SIGMA {\bf 18}, 016, (2022).

\bibitem{APV}
S. Alexandrov, B. Pioline, S. Vandoren,
{\em Self-dual Einstein Spaces, Heavenly Metrics and Twistors},
J. Math. Phys. {\bf 51}, 073510, (2010).

\bibitem{A25}
B. Araneda,
{\em Teukolsky equations, twistor functions, and conformally self-dual spaces},
Lett. Math. Phys. {\bf 115}, 89, (2025).
arXiv:2407.10939.

\bibitem{BBHKMS24}
R. Bittleston, G. Bogna, S. Heuveline, A. Kmec, L. J.  Mason, D. Skinner,
{\em On AdS$_4$ deformations of celestial symmetries},
JHEP {\bf 07}, 010, (2024).
arXiv:2403.18011.

\bibitem{BHS}
R. Bittleston, S. Heuveline, D. Skinner,
{\em The celestial chiral algebra of self-dual gravity on Eguchi-Hanson space},
JHEP {\bf 09}, 008, (2023).

\bibitem{BH26}
G. Bogna, S. Heuveline,
{\em Toward celestial chiral algebras of self-dual black holes},
Phys. Rev. D {\bf 113}, 046010, (2026).

\bibitem{BS21}
T. Bridgeland, I. A. B. Strachan,
{\em Complex hyperk\"ahler structures defined by Donaldson--Thomas invariants},
Lett. Math. Phys. {\bf 111}, 54, (2021).

\bibitem{CP}
A. Chudecki, M. Przanowski,
{\em Killing Symmetries in $\mathcal{H}$-Spaces with $\Lambda$},
J. Math. Phys. {\bf 54}, 102503, (2013).

\bibitem{CS22}
V. Cortés, A. Saha,
{\em Four-dimensional Einstein manifolds with Heisenberg symmetry},
Ann. Mat. Pura Appl. {\bf 201}, 1075--1095, (2022).

\bibitem{DBook}
M. Dunajski,
{\em Solitons, Instantons and Twistors},
Oxford Graduate Texts in Mathematics {\bf 19},
Oxford University Press, (2010).

\bibitem{DFK}
M. Dunajski, E. V. Ferapontov, B. Kruglikov,
{\em On the Einstein-Weyl and conformal self-duality equations},
J. Math. Phys. {\bf 56}, 083501, (2015).

\bibitem{DM00}
M. Dunajski, L. J. Mason,
{\em Hyper-K\"ahler hierarchies and their twistor theory},
Commun. Math. Phys. {\bf 213}, 641--672, (2000).

\bibitem{DM26}
M. Dunajski, T. Moy,
{\em Heavenly equations in de Sitter space},
JHEP {\bf 09}, 247, (2026).

\bibitem{FinPleb}
J. D. Finley III, J. F. Pleba\'nski,
{\em The intrinsic spinorial structure of hyperheavens},
J. Math. Phys. {\bf 17}, 2207--2214, (1976).

\bibitem{GP78}
G. W. Gibbons, C. N. Pope,
{\em $\mathbb{CP}^{2}$ as a gravitational instanton},
Commun. Math. Phys. {\bf 61}, 239--248, (1978).

\bibitem{Hoeg}
M. H\"ogner,
{\em Quaternion-K\"ahler four-manifolds and Przanowski's function},
J. Math. Phys. {\bf 53}, 103517, (2012).

\bibitem{Kim25}
J.-H. Kim,
{\em Single Kerr-Schild metric for Taub-NUT instanton},
Phys. Rev. D {\bf 111}, L021703, (2025).

\bibitem{Kim26}
J.-H. Kim,
{\em Incidence relations for self-dual black holes},
arXiv:2608.07774, (2026).

\bibitem{KL25}
K. Krasnov, A.E. Lipstein,
{\em Pure connection formalism and Pleba\'nski's second heavenly equation},
JHEP {\bf 03}, 152, (2025).

\bibitem{LN}
A. E. Lipstein, S. Nagy,
{\em Self-dual gravity and color-kinematics duality in $AdS_4$},
Phys. Rev. Lett. {\bf 131}, 081501, (2023).

\bibitem{MW}
L. J. Mason, N. M. J. Woodhouse,
{\em Integrability, Self-Duality, and Twistor Theory},
London Mathematical Society Monographs, New Series {\bf 15},
Oxford University Press, (1996).

\bibitem{Miller1}
N. Miller,
{\em Proof of the graviton MHV formula using Pleba\'nski's second heavenly equation},
JHEP {\bf 10}, 146, (2025).

\bibitem{Miller2}
N. Miller,
{\em Spacetime $Lw_{1+\infty}$ Symmetry and Self-Dual Gravity in Pleba\'nski Gauge},
arXiv:2504.07176, (2025).

\bibitem{MoyThesis}
T. Moy,
{\em Joyce Structures and Quaternionic Geometry},
PhD thesis, University of Cambridge, (2026).

\bibitem{N17}
M. Nozawa,
{\em Euclidean supersymmetric solutions with the self-dual Weyl tensor},
Phys. Lett. B {\bf 770}, 166--173, (2017).

\bibitem{Pedersen}
H. Pedersen,
{\em Einstein metrics, spinning top motions and monopoles},
Math. Ann. {\bf 274}, 35--59, (1986).

\bibitem{Penrose}
R. Penrose,
{\em Nonlinear gravitons and curved twistor theory},
Gen. Relativ. Gravit. {\bf 7}, 31--52, (1976).

\bibitem{PR84}
R. Penrose, W. Rindler,
{\em Spinors and Space-Time, Volume 1:
Two-Spinor Calculus and Relativistic Fields},
Cambridge Monographs on Mathematical Physics,
Cambridge University Press, (1984).

\bibitem{PR86}
R. Penrose, W. Rindler,
{\em Spinors and Space-Time, Volume 2:
Spinor and Twistor Methods in Space-Time Geometry},
Cambridge Monographs on Mathematical Physics,
Cambridge University Press, (1986).

\bibitem{Pl76}
J. F. Pleba\'nski,
{\em Some solutions of complex Einstein equations},
J. Math. Phys. {\bf 16}, 2395--2402, (1975).

\bibitem{PrzanKill}
M. Przanowski,
{\em Locally Hermit--Einstein, Self-Dual Gravitational Instantons},
Acta Phys. Polon. B {\bf 14}, 625--627, (1983).

\bibitem{Przan85}
M. Przanowski,
{\em $\bar{\partial}$-K\"ahlerian manifolds and self-dual gravitational instantons},
Phys. Lett. A {\bf 110}, 295--297, (1985).

\bibitem{PrzanHerm}
M. Przanowski,
{\em Killing vector fields in self-dual, Euclidean Einstein
spaces with $\Lambda \ne 0$},
J. Math. Phys. {\bf 32}, 1004--1010, (1991).

\bibitem{Str95}
I. A. B. Strachan,
{\em Integrable systems and curved twistor spaces},
in {\em Further Advances in Twistor Theory, Volume II:
Integrable Systems, Conformal Geometry and Gravitation},
L. J. Mason, L. P. Hughston, P. Z. Kobak (eds.),
Longman, 27--29, (1995).

\bibitem{SW26}
A. Strominger, H. Wei,
{\em Every 3D conformal field theory has an
$\mathcal{L}_{\Lambda}w_{1+\infty}$ symmetry},
Phys. Rev. Lett. {\bf 137}, 221601, (2026).
arXiv:2603.26459.

\bibitem{TZ24}
T. R. Taylor, B. Zhu,
{\em $w_{1+\infty}$ algebra with a cosmological constant
and the celestial sphere},
Phys. Rev. Lett. {\bf 132}, 221602, (2024).
arXiv:2312.00876.

\bibitem{Tod97}
K. P. Tod,
{\em The $SU(\infty)$-Toda field equation and special four-dimensional metrics},
in {\em Geometry and Physics},
J. E. Andersen, J. Dupont, H. Pedersen, A. Swann (eds.),
Lecture Notes in Pure and Applied Mathematics {\bf 184},
Marcel Dekker, New York, 307--312, (1997).

\bibitem{Tod06}
K.P. Tod,
{\em A Note on Riemannian Anti-self-dual Einstein metrics with Symmetry},
arXiv:hep-th/0609071, (2006).

\bibitem{Ward}
R. S. Ward,
{\em Self-dual space-times with cosmological constant},
Commun. Math. Phys. {\bf 78}, 1--17, (1980).

\bibitem{Z02}
K. Zoubos,
{\em Holography and Quaternionic Taub--NUT},
JHEP {\bf 12}, 037, (2002).
arXiv:hep-th/0209235.

\end{thebibliography}
\end{document}